\documentclass[]{aa}
\usepackage{natbib}         % pour A&A
\usepackage{graphicx}
\usepackage[switch]{lineno} % switch=Put the line into the inner margin
\linenumbers\modulolinenumbers[5]
\bibpunct{(}{)}{;}{a}{}{,} % to follow the A&A style

\newcommand{\ind}[1]{_{\mathrm{#1}}}
\newcommand{\diff}{\mathrm{d}}

\def\nombre{218}
\def\nDnu{1180}
\def\nTobs{640}
\def\epsgval{0.5}

\def\Kepler{\emph{Kepler}}
\def\numax{\nu\ind{max}}
\def\nmax{n\ind{max}}
\def\Dnu{\Delta\nu}
\def\nunc{\hat\nu}
\def\dnugr{\delta\nu\ind{rot,g}}
\newcommand\Tg{\Delta\Pi_1}
\newcommand{\tgobs}{\Delta P\ind{obs}}
\newcommand\epsg{\varepsilon\ind{g}}
\newcommand{\per}{P}% BV
\newcommand{\BV}{Brunt-V\"ais\"al\"a}
\newcommand{\NBV}{N\ind{BV}}

\renewcommand{\ng}{n\ind{g}}
\newcommand{\nm}{n\ind{m}}
\newcommand{\np}{n\ind{p}}
\def\fac{\mathcal{N}}
\def\alphag{\alpha_1}
\def\gmmode{g-m mode}
\def\pmmode{p-m mode}
\def\Fcoup{\mathcal{F}}
\def\Frot{\mathcal{R}}
\def\coup{q}
\def\coupm{\langle\coup\rangle}

\def\Teff{T\ind{eff}}
\begin{document}
\title{Probing the core structure and evolution of red giants using
gravity-dominated mixed modes observed with Kepler\thanks{Full
version of Table 1 is only available in electronic form at the CDS
via anonymous ftp to \texttt{cdsarc.u-strasbg.fr (130.79.128.5)}
or via \texttt{http://cdsweb.u-strasbg.fr/cgi-bin/qcat?J/A+A/}}}
\titlerunning{Mixed modes in red giants}

%\author{B. Mosser et al.
\author{B. Mosser\inst{1}\and M.J. Goupil\inst{1}
\and K. Belkacem\inst{1} \and E. Michel\inst{1}
\and D. Stello\inst{2}
\and J.P. Marques\inst{1,3}
\and Y. Elsworth\inst{4}
\and C. Barban\inst{1}
\and P.G. Beck\inst{5}
\and T.R. Bedding\inst{2}
\and J. De Ridder\inst{5}
\and R.A. Garc{\'\i}a\inst{6}
\and S. Hekker\inst{7,4}
\and T. Kallinger\inst{5,8}
\and R. Samadi\inst{1}
\and M.C. Stumpe\inst{9}
\and T. Barclay\inst{10}
\and C.J. Burke\inst{9}
}

\offprints{B. Mosser}

\institute{LESIA - Observatoire de Paris, CNRS, Universit\'e Pierre et Marie Curie, Universit\'e Denis Diderot,
 92195 Meudon cedex, France; \email{benoit.mosser@obspm.fr}
\and Sydney Institute for Astronomy, School of Physics, University of Sydney, NSW 2006, Australia
\and Georg-August-Universit\"at, Institut f\"ur Astrophysik, Friedrich-Hund-Platz 1, D-37077 G\"ottingen Germany
\and School of Physics and Astronomy, University of Birmingham, Edgbaston, Birmingham B15 2TT, United Kingdom
\and Instituut voor Sterrenkunde, K. U. Leuven, Celestijnenlaan 200D, 3001 Leuven, Belgium
\and Laboratoire AIM, CEA/DSM – CNRS - Universit\'e Paris Diderot – IRFU/SAp, 91191 Gif-sur-Yvette Cedex, France
\and Astronomical Institute `Anton Pannekoek', University of Amsterdam, Science Park 904, 1098 XH Amsterdam, The Netherlands
\and Institute for Astronomy (IfA), University of Vienna, T\"urkenschanzstrasse 17, 1180 Vienna, Austria
%\and High Altitude Observatory, NCAR, P.O. Box 3000, Boulder, CO 80307, USA
%\and Institut d'Astrophysique Spatiale, UMR 8617, Universit\'e Paris XI, B\^atiment 121, 91405 Orsay Cedex, France
%\and Institut d'Astrophysique et de G\'eophysique, Universit\'e de Li\`ege, All\'ee du 6 Ao\^ut, 17 B-4000 Li\`ege, Belgium
%\and Orbital Sciences Corporation/NASA Ames Research Center, Moffett Field, CA 94035, USA
\and SETI Institute/NASA Ames Research Center, Moffett Field, CA
94035, USA \and Bay Area Environmental Research Inst./NASA Ames
Research Center, Moffett Field, CA 94035, USA }
%\date{Submitted to A\&A}

\abstract{There are now more than 22 months of long-cadence data
available for thousands of red giants observed with the \Kepler\
space mission. Consequently, we are able to clearly resolve fine
details in their oscillation spectra and see many components of
the mixed modes that probe the stellar core.}
{We report for the first time a parametric fit to the pattern of
the $\ell=1$ mixed modes in red giants, which is a powerful tool
to identify gravity-dominated mixed modes. With these modes, which
share the characteristics of pressure and gravity modes, we are
able to probe directly the helium core and the surrounding shell
where hydrogen is burning.}
{We propose two ways for describing the so-called mode bumping
that affects the frequencies of the mixed modes. Firstly, a
phenomenological approach is used to describe the main features of
the mode bumping. Alternatively, a quasi-asymptotic mixed-mode
relation provides a powerful link between seismic observations and
the stellar interior structure. We used period \'echelle diagrams
to emphasize the detection of the gravity-dominated mixed modes.
}%
{The asymptotic relation for mixed modes is confirmed. It allows
us to measure the gravity-mode period spacings in more than two
hundred red giant stars. The identification of the
gravity-dominated mixed modes allows us to complete the
identification of all major peaks in a red giant oscillation
spectrum, with significant consequences for the true
identification of $\ell=3$ modes, of $\ell=2$ mixed modes, for the
mode widths and amplitudes, and for the $\ell=1$ rotational
splittings.
}%
{The accurate measurement of the gravity-mode period spacing
provides an effective probe of the inner, g-mode cavity. The
derived value of the coupling coefficient between the cavities is
different for red giant branch and clump stars. This provides a
probe of the hydrogen-shell burning region that surrounds the
helium core. Core contraction as red giants ascend the red giant
branch can be explored using the variation of the gravity-mode
spacing as a function of the mean large separation.}

\keywords{Stars: oscillations - Stars: interiors - Stars:
evolution - Stars: late-type - Methods: data analysis -
Asteroseismology}

\maketitle %\voffset = 1.5cm
%________________________________________________________________

\section{Introduction\label{introduction}}

The NASA \Kepler\ mission provides us with thousands of
high-precision photometric light curves of red giant stars
\citep{2010Sci...327..977B,2010PASP..122..131G,2010ApJ...713L..79K}.
This combination of long duration and uninterrupted data allows us
to study the properties of the red giant oscillation spectra.
Studying these oscillations is providing a clear insight into the
stellar structure, and was already initiated by CoRoT observations
\citep{2009Natur.459..398D}. Owing to the unprecedented quality of
the CoRoT and \Kepler\ data, many ensemble analyses of red giant
oscillations have been performed
\citep[e.g.][]{2009A&A...506..465H,2010ApJ...713L.176B,
2010A&A...517A..22M,2010ApJ...723.1607H,2010A&A...522A...1K}, as
well as a few studies dedicated to individual objects
\citep{2010A&A...509A..73C,2010A&A...520L...6M,2011ApJ...742..120J,
2011MNRAS.415.3783D,2012A&A...538A..73B}. A comprehensive review
of these recent observations is given by
\cite{2011arXiv1107.1723B}.

Up to now, much of the information has come from the pressure-mode
pattern, which probes primarily the external region of the stars.
A direct probe of the core regions is provided by the mixed modes,
which propagate as pressure waves in the convective envelope, and
as gravity waves in the radiative region of the core
\citep[e.g.][]{1977A&A....58...41A}. Mixed modes, theoretically
described in previous work \citep[e.g.][and references
therein]{2001MNRAS.328..601D,2004SoPh..220..137C,2009A&A...506...57D},
were reported in red giants by \cite{2010ApJ...713L.176B}, who
attributed the broadening of the $\ell=1$ ridge to the presence of
multiple $\ell = 1$ peaks whose frequencies are shifted by avoided
crossings. Their period spacings were first measured by
\cite{2011Sci...332..205B}. \cite{2011Natur.471..608B} and
\cite{2011A&A...532A..86M} have shown the capability of these
modes to measure the evolutionary status of red giants, with a
clear difference between stars ascending the red giant branch
(RGB) and clump stars.

\cite{2012A&A...537A..30M} showed that the mixed-mode pattern
sometimes has a very large extent, with $\ell=1$ mixed modes
located very far from the expected position of the $\ell=1$ pure
pressure modes. In this paper, we study these gravity-dominated
mixed modes, hereafter called \gmmode s. In contrast to the
pressure-dominated mixed modes, or \pmmode s, they have the
characteristics of gravity modes: their period spacing, hereafter
denoted $\Tg$, is close to the spacing of pure asymptotic g modes
and they have much narrower widths. In red giant spectra, there
are only a few dipole ($\ell=1$) \pmmode s per radial order, but
several \gmmode s.

The analysis by \cite{2011Natur.471..608B} has shown that the
coupling between gravity and pressure waves induces a mode bumping
that decreases the spacing between adjacent mixed modes and
complicates the determination of the gravity-mode period spacing.
A simple adhoc model presented in this work shows how this bumping
is related to the avoided crossings between the g modes and the
pure p mode that one would observe in absence of any coupling. It
induces a local dip in the mixed-mode period spacing. The measured
value of the period spacing, $\tgobs$, between bumped mixed modes
is significantly lower than the period spacing $\Tg$ and does not
give access to it. With the identification of mixed modes in a
wider frequency range, we may now establish a direct measure of
$\Tg$. This measurement is of prime importance for addressing the
physical conditions in the stellar core.

\begin{figure*}
\includegraphics[width=17cm]{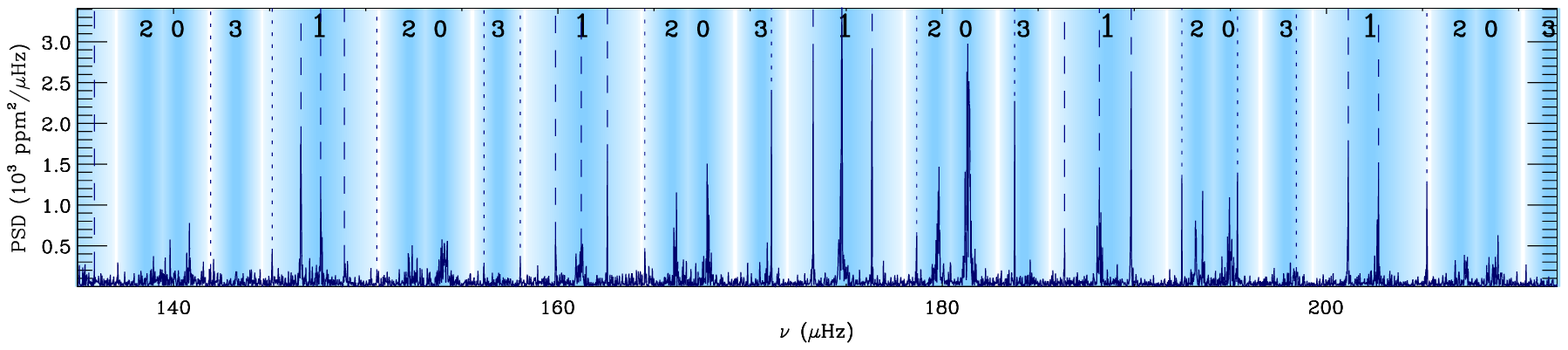}
\includegraphics[width=17cm]{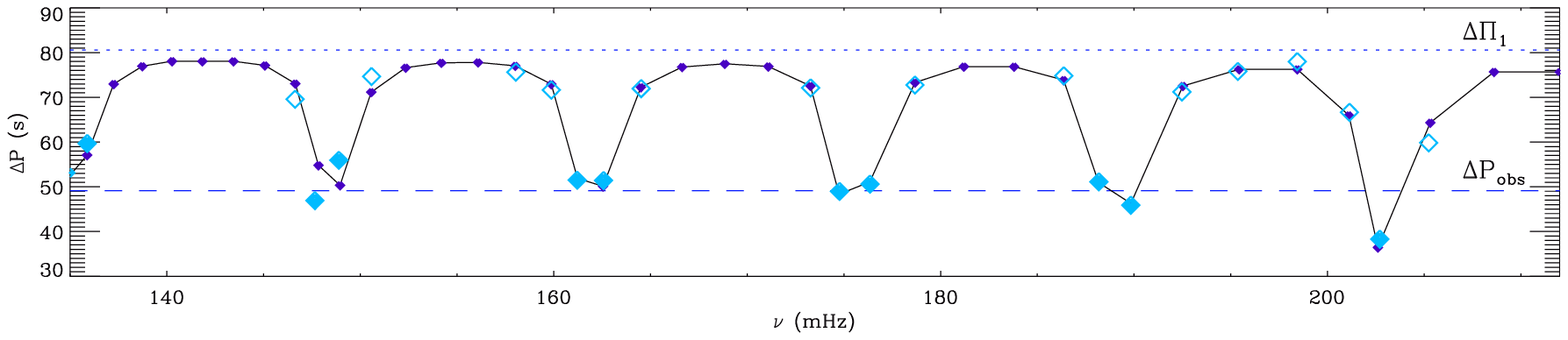}
\caption{{\sl Top:} Power density spectrum of the star KIC
9882316, with superimposed mode identification provided by the red
giant oscillation universal pattern. Dashed (dotted) lines
indicates the position of the peaks a priori (a posteriori)
identified as $\ell=1$ mixed modes. P-m modes are located close to
the positions marked by 1. {\sl Bottom:} Period spacings
$\Delta\per= 1/\nu_{\nm,1} - 1/\nu_{\nm+1,1}$ between adjacent
mixed modes, as a function of the frequency. The full line and
small filled diamonds correspond to the convolution model defined
by Eqs.~\ref{convol} and \ref{couplingF}; the g-mode spacing $\Tg$
is indicated by the dotted horizontal line; the dashed line
indicates the observed spacing, $\tgobs$, which is affected by the
mode bumping, measured with the method presented by
\cite{2011A&A...532A..86M}. The large diamonds show the spacing
between two observed consecutive modes; full symbols, near the
dashed line, correspond to the measurements derived from the modes
a priori identified as $\ell=1$ mixed modes; open symbols, just
below the dotted line, are obtained from the peaks identified a
posteriori as mixed modes. \label{example}}
\end{figure*}

In Sect.~\ref{obs}, we briefly present the \Kepler\ observations
used in this work and explain how \gmmode s are identified. Two
ways to model the \gmmode\ pattern are proposed in
Sect.~\ref{mixedmodes}. Firstly, a simple adhoc model allows us to
identify and fit the mixed-mode pattern, and then to relate the
mode bumping to the pure p and g mode patterns in red giants.
Secondly, we adopt a more physical approach based on the
asymptotic formalism for mixed modes developed by
\cite{1979PASJ...31...87S} and \cite{1989nos..book.....U}, and
specifically adapted to subgiants and red giants by Goupil et al.
(in prep.).

The quality of the asymptotic fit to the data is investigated in
Sect.~\ref{method}. Inferences of this work are presented in
Sect.~\ref{discussion}, where we show that \gmmode s probe
simultaneously the physical conditions in the stellar core and the
interface between the core and the envelope, where hydrogen burns
in a shell.

\section{Observations\label{obs}}

\subsection{Data}

The data used are mainly \Kepler\ long-cadence data that have been
described in previous papers \citep[e.g.][and references
therein]{2010ApJ...713L..87J,2010ApJ...713L.176B,2010A&A...522A...1K,2010ApJ...723.1607H,
2011ApJ...741..119M,2012A&A...537A..30M}. Original light curves
were processed and corrected according to the procedure of
\cite{2011MNRAS.414L...6G}. We analyzed the stars with the highest
signal-to-noise ratio spectra obtained from 690-day time series up
to the \Kepler\ quarter Q8. The frequency resolution of 16.8\,nHz
allowed us to measure the global seismic parameters $\Dnu$ (large
separation between consecutive radial orders) and $\numax$
(central frequency of the oscillation power excess) in \nDnu\
targets using the envelope autocorrelation method
\citep{2009A&A...508..877M}. Results provided by this method agree
closely with other methods, as shown by
\cite{2011A&A...525A.131H}.

The evolutionary status of the red giants was determined for
\nTobs\ stars from $\tgobs$, measured using the automated method
developed by \cite{2011A&A...532A..86M}. This method requires the
precise determination of the expected location of the theoretical
$\ell=1$ pure p modes. As the power-of-power method described by
\cite{2011Natur.471..608B}, it is mostly sensitive to the bumped
spacing $\tgobs$  and is therefore only a lower limit for the true
g-mode spacing $\Tg$.

Since we deal in this work with period spacing measurements, it is
convenient to translate the frequency resolution into period
resolution. At 20, 40, 80, and 160\,$\mu$Hz, this corresponds to
about 49, 12, 3, and 0.8\,s, respectively. This rapid variation
(as $\nu^{-2}$) shows that our ability to measure period spacings
declines strongly at low frequency.

The oscillation spectra of long-cadence data have a Nyquist
frequency of 283.5\,$\mu$Hz. To include stars from the lower part
of the RGB in our analysis, we must turn to the \Kepler\
short-cadence data (58.8 s instead of 29.4 min), whose Nyquist
frequency is much higher. However, since we are only interested in
the red giants and not in subgiants or main-sequence stars, we
must add an upper limit to the frequency range that we use. It is
easy to exclude the main-sequence stars because they are at much
higher frequencies than we consider and do not show mixed modes.
Excluding the subgiants is more problematic. To guide us in these
choices we searched published analyses of individual CoRoT and
\Kepler\ giants and subgiants
\citep[see][]{2011MNRAS.415.3783D,2011ApJ...742..120J,2010A&A...515A..87D,2011ApJ...733...95M,2011A&A...534A...6C}.
Based on these results, we derived an upper limit on the large
separation of $\Dnu = 40\,\mu$Hz. Note that this is a conservative
limit; we neither claim that all red giants have $\Dnu \le
40\,\mu$Hz, nor that all stars with $\Dnu \le 40\,\mu$Hz are red
giants. From the analysis of the red giant radii
\citep{2011ApJ...738L..28V} and of the near-surface offsets
\citep{2011ApJ...742L...3W}, we verified that the limit value at
$40\,\mu$Hz is adequate for our work.

Our sample includes the two red giants with long-duration data
sets from the publications listed above, plus a selection from the
one-month short-cadence data survey data
\citep{2010ApJ...713L.169C}. These comprise 15 stars with high
signal-to-noise oscillation spectra and $\Dnu$ in the range [20,
40\,$\mu$Hz], to complement the long-cadence data. The
signal-to-noise ratio is defined from the envelope autocorrelation
function \citep{2009A&A...508..877M}. In making the selection, we
were also mindful of the need to include stars with a wide range
of masses, the mass being estimated from the asteroseismic scaling
relation, as in \cite{2010A&A...517A..22M}, and using the
effective temperature given in the \Kepler\ Input Catalog
\citep{2011AJ....142..112B}.

As shown below in Fig.~\ref{spacings}, the selected short-cadence
stars in the less evolved region  of the RGB show properties that
extend and are consistent with the  analysis primarily based on
long-cadence data.

\subsection{Detection of $\ell=1$ \gmmode s\label{gmmode}}

Red giant oscillation observations have revealed many extra modes
that are not explained by the p-mode asymptotic pattern
\citep{2010ApJ...713L.176B,2010A&A...517A..22M}. These modes have
been identified as $\ell=1$ mixed modes
\citep{2011Sci...332..205B}. However, further work showed that
some giants, although not all, show very many mixed modes. A
careful examination of a few cases has revealed that these modes
are present everywhere in the spectrum, not just close to the
expected location of the pure p modes. A representative example is
shown in Fig.~\ref{example}. Dozens of similar cases have been
found, and \nombre\ of them are considered in this paper.

A multi-stage process was used to identify the $\ell=1$ \gmmode s
that lie far from the theoretical location of pure p modes. First,
the determination of the background
\citep{2011ApJ...741..119M,2012A&A...537A..30M} allowed us to
identify peaks that can be reliably attributed to oscillations.
Then, we used the so-called universal pattern to locate the
short-lived radial and non-radial pressure modes very precisely
\citep{2011A&A...525L...9M}. The universal pattern provides a
global description of the red giant oscillation pattern.
Comparison with a local description is given by \cite{kallinger}.

According to \cite{2009A&A...506...57D}, mixed modes should have
noticeable amplitudes only close to the location of the pure p
modes. However, there is evidence from the data that mixed modes
are observed throughout the spectrum. Given the relative positions
of the g- and p-mode cavities for modes of different degree, the
$\ell=1$ mixed modes are most likely to be visible, which is what
we assume here. Therefore, unassigned peaks in the spectrum were
considered as candidates for being dipole \gmmode s and were used
to construct a period-spacing diagram (Fig.~\ref{example}). The
threshold level used for excluding peaks caused by noise was set
empirically at eight times the background value, such that the
probability of including a spurious noise
peak was less than 1/100. %moins que 1/3 pour une moyenne de 30 pics.
If rotational splitting was present \citep{2012Natur.481...55B} we
kept only the central $m=0$ component.

By focusing on red giant spectra with very many \gmmode s
(Fig.~\ref{example}), we can address the full identification of
the $\ell=1$ mixed-mode pattern. The measurement of $\Tg$, rather
than only $\tgobs$, requires a dedicated modeling as presented
below.

\section{Parameterization of mixed modes\label{mixedmodes}}

%%%%%%%%%%%%%%%%%%%%%%%%%%%%%%%%%%%%%%%%%%%%%%%%%%%%%%%%%%

Mixed modes result from the coupling of p and g waves. We have
adopted the description of the universal oscillation pattern of
red giants for the p modes \citep{2011A&A...525L...9M}. The pure
p-mode eigenfrequency pattern can then be expressed as
\begin{equation}\label{tassoulp}
\nu_{\np,\ell} = \left(\np +{\ell \over 2}+\varepsilon (\Dnu ) -
d_{0\ell} (\Dnu ) + {\alpha \over 2}[ \np- \nmax ]^2 \right) \Dnu,
\end{equation}
where $\Dnu$ is the large separation, $\np$ is the p-mode radial
order, $\ell$ is the angular degree, $\varepsilon$ is the phase
offset (which is a function of $\Dnu$), $d_{0\ell}$ accounts for
the so-called small separation, and $\nmax= \numax / \Dnu$. The
constant $\alpha$ represents the mean curvature of the p-mode
oscillation pattern and has a value of about 0.008, which is
derived from the detailed analysis of the radial-oscillation
pattern with the method presented by \cite{2010AN....331..944M}.

We next consider the asymptotic development of pure gravity dipole
modes derived by \cite{1980ApJS...43..469T}. To first order, the
periods follow
\begin{equation}
\per_{\ng, \ell=1} = (|\ng| + \alphag)\;  \Tg \ , \label{tassoulg}
\end{equation}
where $\ng$ is the gravity radial order, $\alphag$ is a constant,
and $\Tg$ is the period spacing of dipole g modes. Conventionally,
the radial order $\ng$ is defined as a negative integer, in
contrast to the positive $\np$ order, so that the g-mode
eigenfrequencies are increasing with increasing order. In the
asymptotic limit \citep[e.g.][]{1977AcA....27...95D}, the period
$\Tg$ is related to the \BV\ frequency $\NBV$ according to
\begin{equation}
\Tg = % {\Delta\mathcal{T\ind{g}} \over \sqrt{2}} = %
{2\pi^2 \over \sqrt{2}} \ \left[ \int_{\mathrm{core}} \!\!{{\NBV
\over r}\; \diff r } \right]^{-1}.
 \label{def_period}
\end{equation}
Measuring $\Tg$ gives access to an integral of the \BV\ frequency
weighted by the inverse of the radius. In red giants, the high
density reached in the core gives a high value of $\NBV$. The
increase of the core density expected when the star evolves on the
RGB results in a decrease of the gravity spacing. When helium
ignition occurs in the core, the energy release gives rise to a
convective region where $\NBV$ vanishes. This should translate
into an increase of $\Tg$ \citep{2011arXiv1106.5946C}. Hence, we
can directly probe the stellar core through the measurement of the
\BV\ frequency.

\subsection{Fitting procedure: an empirical approach}

To estimate $\Tg$, we first developed an empirical approach for
modeling the mode bumping. \cite{2010Ap&SS.328..259D} carried out
an elegant analysis based on coupled oscillators. Here, we adopt
an alternative approach relying on asymptotic relations. In the
absence of coupling, the eigenfrequency pattern would simply be
the combination of the p- and g-mode asymptotic patterns. We
denote this pattern by $\nunc (k)$, restricted to dipole modes,
with the index $k$ enumerating the eigenfrequencies. In the
$\Dnu$-wide interval centered on $\numax$, according to
Eqs.~\ref{tassoulp} and \ref{tassoulg} we have one dipole p mode
and $\fac \simeq \Dnu\,\Tg^{-1}\numax^{-2} $ g modes.

To enumerate the mixed modes, we need to introduce a mixed-mode
index. We define it as $\nm = \ng + \np$, with $\ng$ and $\np$
being the gravity and pressure radial orders. This definition,
with negative values of $\ng$, hence of $\nm$, provides an
accurate and continuous numbering of the mixed modes. However, it
is not equal to the mixed-mode radial order, which indicates the
number of radial nodes in the wavefunction and is given by $|\ng|
+ \np$.

A non-negligible coupling is observed (Fig.~\ref{example}), so
that we have no direct access to the uncoupled p and g
eigenfrequencies $\nunc$. We found that the mixed-mode pattern can
be reproduced by a redistribution of the p and g eigenfrequencies
according to the convolution of the unperturbed p-g pattern with a
coupling function $\Fcoup$. This convolution model, which is easy
to implement numerically, reproduces the effect of the coupling:
it simply redistributes the eigenfrequency differences, with the
signature of the theoretical pure p modes expressed by the mode
bumping. Instead of having $\fac$ gravity modes at $\numax$ in a
$\Dnu$-wide interval, there are $\fac + 1$ mixed modes, with modes
bumped in the \pmmode\ region centered on the theoretical position
of the uncoupled p mode. In practice, the eigenfrequency
differences of the uncoupled pattern are redistributed according
to
\begin{equation}\label{convol}
    \nu (\nm) - \nu (\nm-1) = \sum_{k= \nm-\fac}^{\nm+\fac}
\bigl( \nunc(k) - \nunc(k-1) \bigr)\
    \Fcoup_{\nm} (\nunc(k))
    .
\end{equation}
For clarity, we used $\nu (\nm) \equiv \nu_{\nm,\ell=1}$. We
tested different coupling functions and found the best fit with a
Lorentzian:
\begin{equation}\label{couplingF}
    \Fcoup_{\nm} (\nunc) = { 1 \over 1 + \displaystyle{
    \left({\nunc - \nunc(\nm) \over  C\Dnu} \right)^2}}\ ,
\end{equation}
with the coupling factor $C$ , of about 0.2. In
Sect.~\ref{development}, we justify the use of this empirical
function. Importantly, the fit of the \gmmode\ pattern allows us
to determine $\Tg$.

Figure~\ref{example} (bottom) shows the period differences
$\Delta\per = 1/\nu_{\nm,1} - 1/\nu_{\nm+1,1}$ between adjacent
mixed modes as a function of their frequency. All observed period
spacings are smaller than the asymptotic g-mode spacing $\Tg$. The
fit to all observed mixed modes with the convolved frequencies
allows us to identify almost all significant peaks in the power
density spectrum. It then provides a measure of the gravity period
$\Tg$, whereas the fit to the \pmmode s is only able to give
$\tgobs$. While \cite{2011arXiv1107.1311S} found that two
parameters were needed to fit the width and depth of the bumping
of large series of stellar models, we note that a single parameter
$C$ is enough for modeling the mode bumping in the stars presented
here, since the width and the depth of the bumping are correlated.
This is because the mode bumping has to relate the coupling of one
p mode per $\Dnu$-wide frequency range: a low depth implies a
large width, and vice versa.

Finally, the agreement between the observations and the model
allows us to enlarge the set of peaks identified as \gmmode s in
Sect.~\ref{gmmode}. Peaks with a height five times the background
can be assigned to the \gmmode\ pattern when there is close
agreement with the model.

The convolution model provides a very precise but only empirical
fit. Therefore, we have also developed a more physical approach.

\begin{figure*}
\includegraphics[width=3.61cm]{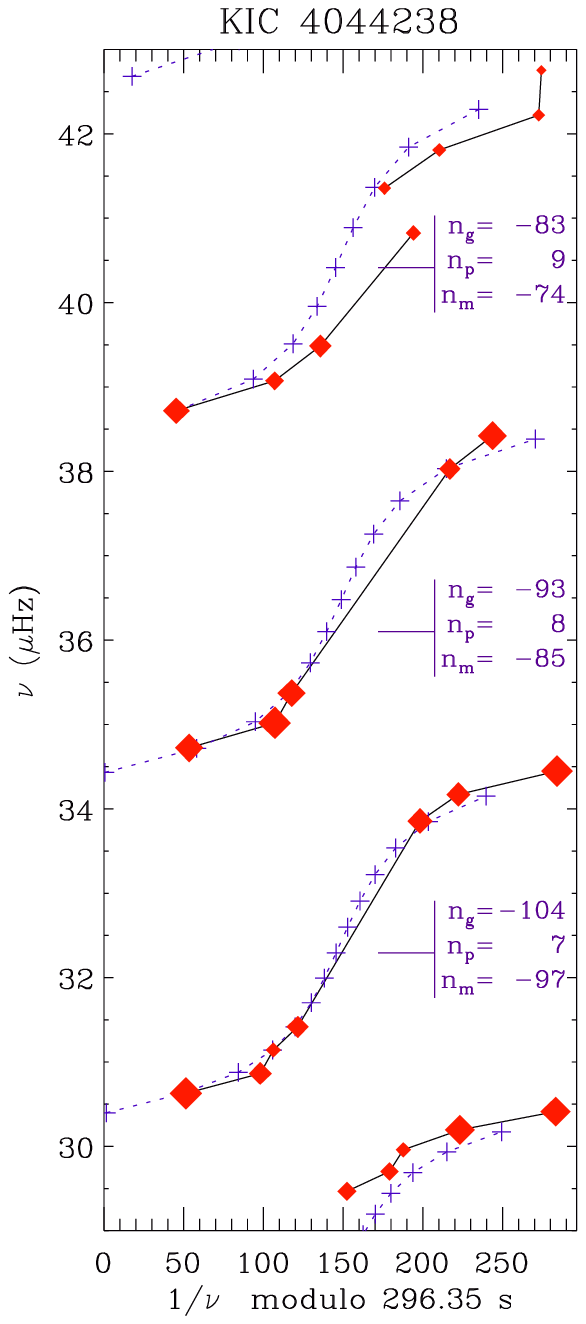}
\includegraphics[width=3.61cm]{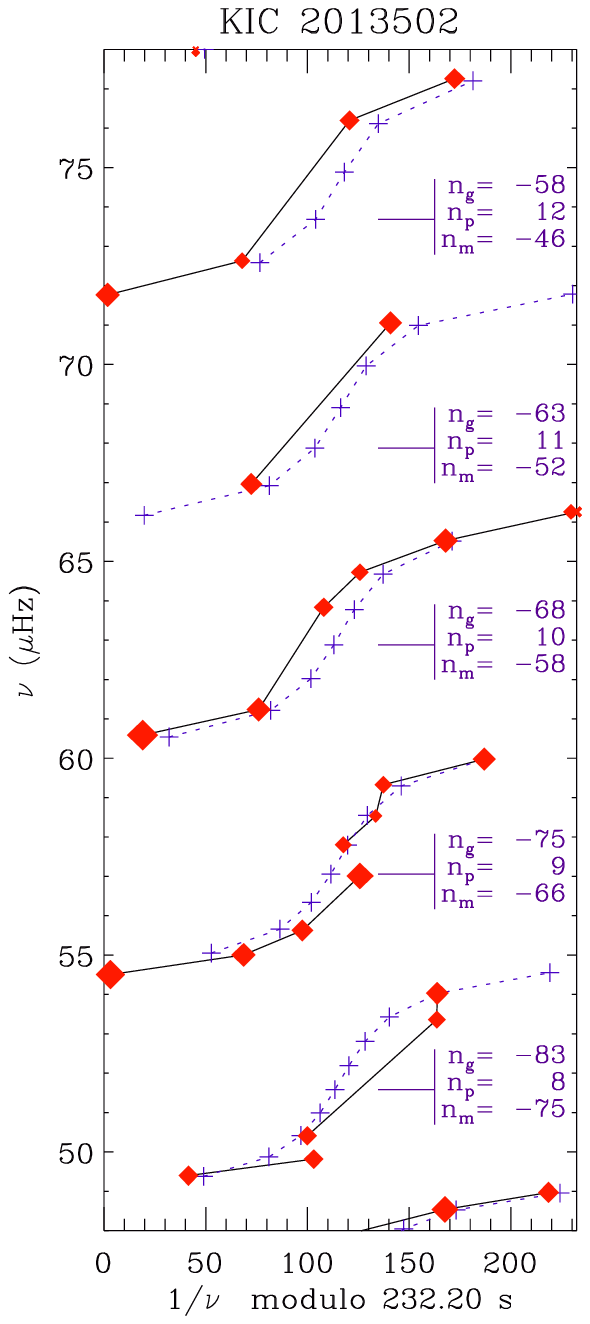}
\includegraphics[width=3.61cm]{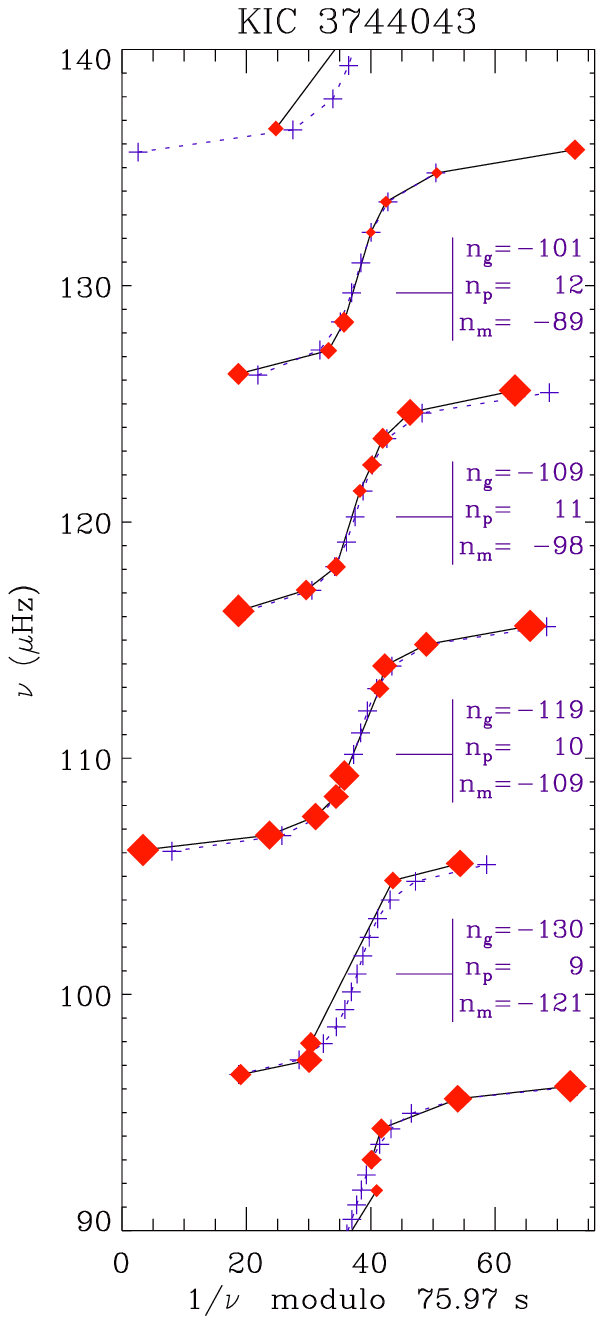}
\includegraphics[width=3.61cm]{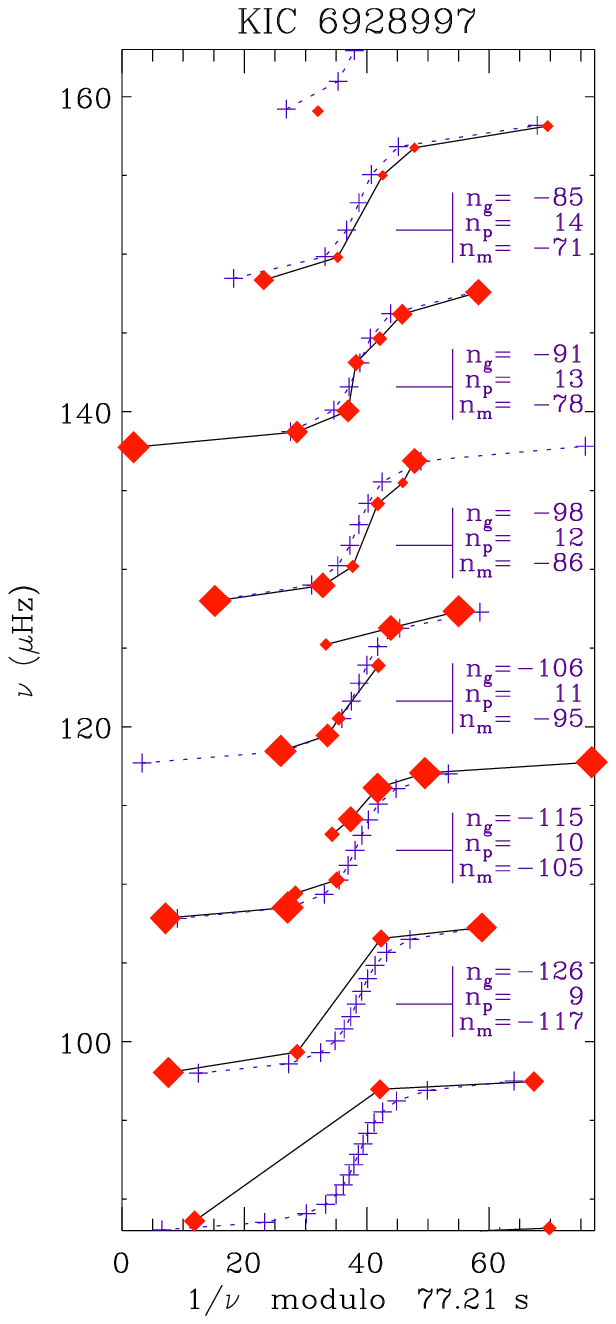}
\includegraphics[width=3.61cm]{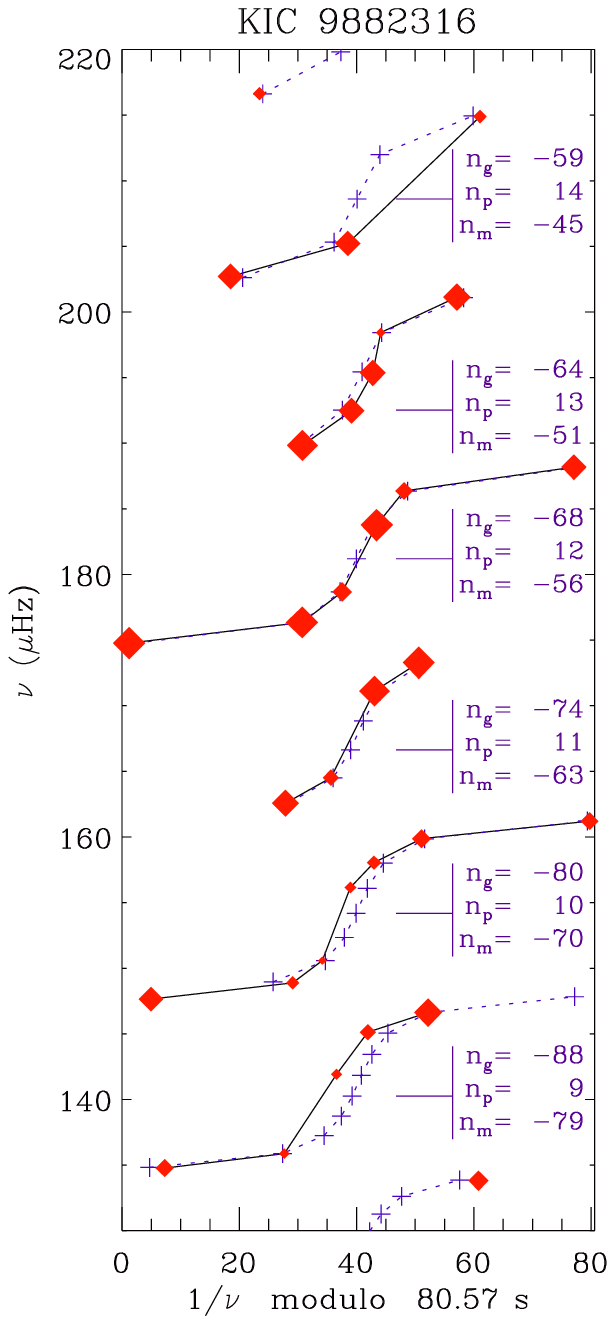}
\caption{Period \'echelle diagrams of five representative stars,
two in the clump (two left-most panels) and three on the RGB,
%stars KIC 4044238, 2013502  , 3744043 and 9882316,
sorted by increasing $\Dnu$. The x-axis shows the period modulo of
the gravity spacing $\Tg$; the y-axis is the frequency. Diamonds
indicate the observed modes, with a size proportional to the mode
height. Dashed lines and crosses correspond to the asymptotic fit.
Pressure and gravity orders ($\np$, $\ng$, respectively) and
mixed-mode index $\nm$ are given for the mixed modes located near
the pressure radial modes. The most p mode-like part of the
pattern is located in the middle of the range, at $1/\nu = \Tg/2\
(\hbox{modulo } \Tg)$. \label{gechelle}}
\end{figure*}

\subsection{Asymptotic development\label{development}}

\cite{1979PASJ...31...87S} and \cite{1989nos..book.....U}
provided an asymptotic relation for p-g mixed modes. In this
framework, eigenfrequencies are derived from an implicit equation
relating the coupling of the p and g waves \citep[][their Eq.
16.50]{1989nos..book.....U}:
\begin{equation}
%\tan \pi {\nu \over \Dnu}  = \coup\ \tan \pi{1 \over \Tg \nu}  ,
\tan \theta\ind{p}  = \coup\ \tan \theta\ind{g}  ,
\label{implicite}
\end{equation}
where $\theta\ind{p}$ and $\theta\ind{g}$ are the p- and g-wave
phases. The dimensionless coefficient $q$ measures the level of
mixture of the p and g phases: $q=0$ is equivalent to no coupling,
and maximum coupling occurs for $q=1$. According to the
first-order development in \cite{1989nos..book.....U}, $q$ is
supposed to be in the range [0, 1/4]. A similar expression was
found by \cite{1992ApJS...80..369B}, but for the coupling of g
waves trapped in two different \BV\ cavities of ZZ Ceti stars.

Equation \ref{implicite} supposes that the asymptotic mixed-mode
relation closely follows the asymptotic relations of p and g modes
(Eqs.~\ref{tassoulp} and \ref{tassoulg}). Owing to the complexity
of the coupling, the asymptotic relation has no explicit
expression. However, following \cite{1989nos..book.....U}, we
consider the p and g phases of Eq.~\ref{implicite}:
\begin{eqnarray}
% \nonumber to remove numbering (before each equation)
  \theta\ind{p} &=& \pi {\nu \over \Dnu} , \\
  \theta\ind{g} &=& \pi {1 \over \nu \Tg}.
\end{eqnarray}
We then introduce the uncoupled solutions of the p modes in
Eq.~\ref{implicite} and express it for the dipole mixed modes
coupled to the pure p mode $\nu_{\np,\ell=1}$ (Eq.~\ref{tassoulp})
as
\begin{equation}
\nu = \nu_{\np,\ell=1} + {\Dnu \over \pi} \arctan %
\left[%
 \coup \tan \pi \left( {1 \over \Tg \nu} -
 \epsg
\right) \right] . \label{implicite2}
\end{equation}
In practice, we assume that all dipole modes in the range
$[\nu_{\np,\ell=1} -\Delta\nu/2 \ ; \ \nu_{\np,\ell=1}
+\Delta\nu/2]$ are coupled with $\nu_{\np,\ell=1}$. The constant
$\epsg$ in Eq.~\ref{implicite2}, already present in
\cite{1992ApJS...80..369B} for the coupling of g waves in two
distinct cavities, derived from Goupil et al. (in prep.) for the
coupling of p and waves, ensures that we obtain \gmmode\ periods
close to $(\ng+1/2+\epsg)\; \Tg$, as expected when the coupling is
weak.

For each pressure radial order $\np$, one derives from
Eq.~\ref{implicite2} $\fac+1$ solutions, with $\fac \simeq
\Dnu\,\Tg^{-1}\numax^{-2}$ as defined previously. The value of
$\Tg$ is then derived from a least-squares fit of the observed
values, defined as in Sect.~\ref{gmmode}, to the asymptotic
solution. Initially, the coupling factor was fixed to be its mean
value $\coupm$, which depends on the evolutionary status. Once
$\Tg$ was determined, the value of the coupling factor was then
iterated, with $\Tg$ being fixed. A stable solution for $\Tg$ and
$\coup$ was found after only a few iterations.

The qualitative agreement of the fit to the mixed-mode asymptotic
relation can be shown in period \'echelle diagrams
(Fig.~\ref{gechelle}). In the classical \'echelle diagram, a
convenient plot of the p-mode pattern, the x-axis is defined as
the frequency modulo of the large separation $\Dnu$. Here, to
represent mixed modes with periods close the g-mode period
pattern, the x-axis of the period \'echelle diagram is defined as
the period modulo $\Tg$ \citep{2011Natur.471..608B}. This
representation is also used for g modes in dense stars
\citep[e.g.][]{2011ApJ...740L..47P}. The absolute position of the
pattern in the period \'echelle diagram depends on the unknown
term $\epsg$, which is supposed to be small. We note that the best
fits gives $1/2 + \epsg\ne 0\ (\hbox{modulo }1)$. Since the fits
do not indicate any trend in $\epsg$, and since its determination
can only be made in modulo 1, we have chosen to fix its value. For
simplicity, we considered $\epsg = 0$, as implicitly assumed by
\cite{2011Natur.471..608B}. The S-shape observed in the period
\'echelle diagram, with one S-pattern per $\Dnu$-wide interval, is
the signature of a coupling with a coupling coefficient close to
$\coupm$. A lower coupling coefficient would correspond to steeper
central segments, whereas stronger coupling would correspond to
more inclined segments.

The asymptotic solutions were compared with those of the empirical
convolution model and they cannot be distinguished. This close
agreement follows from the fact that the Lorentzian form
introduced in Eq.~\ref{couplingF} is the derivative of the
$\arctan$ function, which appears in the asymptotic expression for
the coupling. Strictly speaking, our solutions are only
quasi-asymptotic, since we use in Eq.~\ref{implicite2} the
pressure mode pattern described by Eq.~\ref{tassoulp}, which is
not purely asymptotic. However, for simplicity, we refer to it as
asymptotic. We also note that any a priori dipole pure p-mode
pattern, such as that derived from a precise fit to the radial
modes, can be used for the p-mode frequencies $\nu_{\np, \ell=1}$
in Eq.~\ref{implicite2}.

In the next sections, we examine the results from the fitting to
the observed mixed-mode pattern with the asymptotic relation, and
we use it to derive important information on the red giant
oscillation spectra.

\begin{figure*}
\includegraphics[width=16cm]{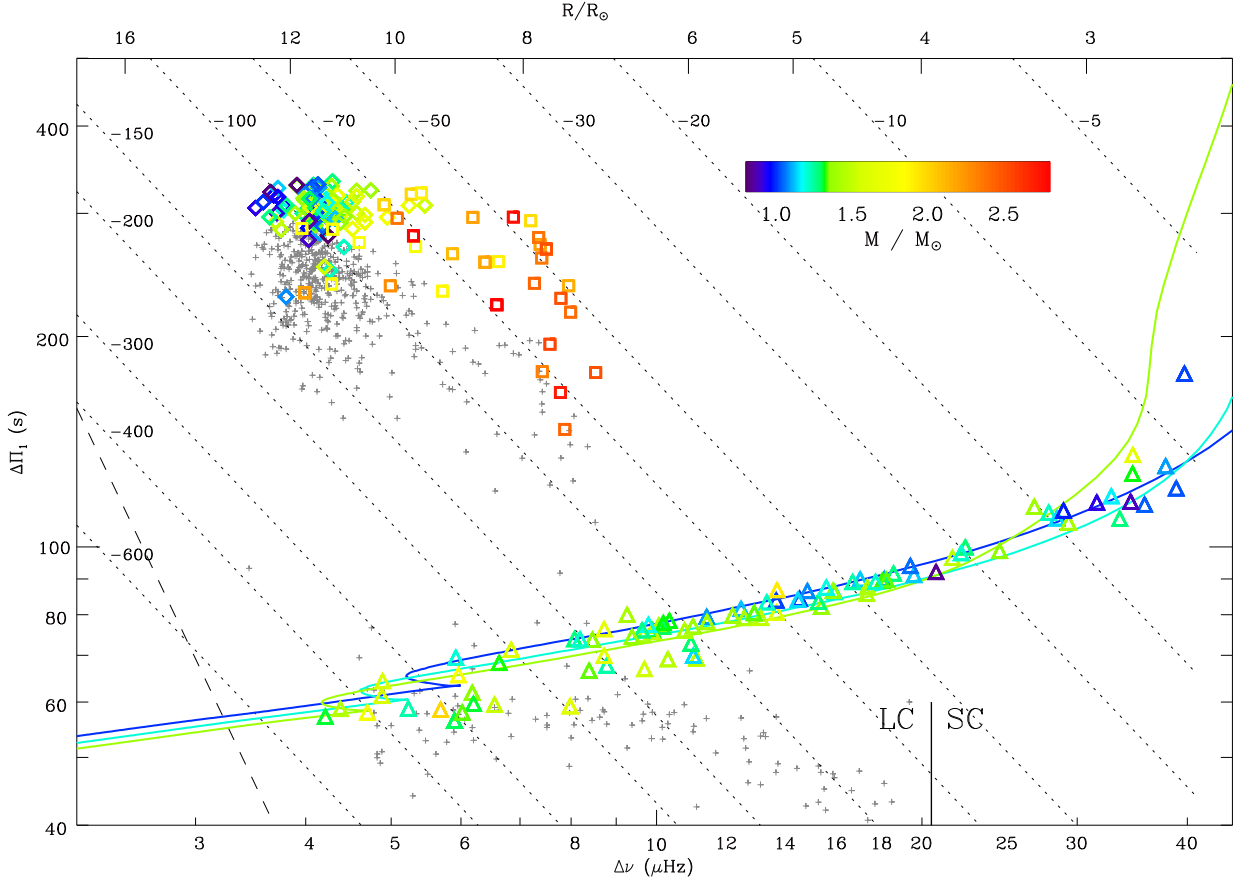}
\caption{Gravity-mode period spacing $\Tg$ as a function of the
pressure-mode large frequency spacing $\Dnu$. Long-cadence data
(LC) have $\Dnu\le 20.4\,\mu$Hz. RGB stars are indicated by
triangles; clump stars by diamonds; secondary clump stars by
squares. Uncertainties in both parameters are smaller than the
symbol size. The seismic estimate of the mass is given by the
color code. Small gray crosses indicate the bumped periods
$\tgobs$ measured by \cite{2011A&A...532A..86M}. Dotted lines are
$\ng$ isolines. The dashed line in the lower left corner indicates
the formal frequency resolution limit. The upper x-axis gives an
estimate of the stellar radius for a star whose $\numax$ is
related to $\Dnu$ according to the mean scaling relation $\numax =
(\Dnu/0.28)^{1.33}$ (both frequencies in $\mu$Hz). The solid
colored lines correspond to a grid of stellar models with masses
of 1, 1.2 and $1.4\, M_\odot$, from the ZAMS to the tip of the
RGB. \label{spacings}}
\end{figure*}

\section{Mixed modes and mode bumping\label{method}}

\subsection{Gravity spacings}

Thanks to the \gmmode s, the fit to the mixed-mode pattern
provides us with the measurement of the gravity spacing $\Tg$. The
number of stars with sufficient \gmmode s is limited to \nombre\
red giants. Measuring $\Tg$ precisely helps us to improve the
criteria for distinguishing the evolutionary status of the stars,
namely that RGB and clump stars clearly show different
distributions, as already observed for the bumped spacings
\citep{2011Natur.471..608B,2011A&A...532A..86M}. A $\Dnu$ --
$\tgobs$ diagram is very useful for distinguishing the
evolutionary status of the stars. We plotted our current results
using such a diagram, but with $\Tg$ instead of $\tgobs$
(Fig.~\ref{spacings}). This diagram allows us to emphasize the
difference between RGB giants, which are burning hydrogen in a
shell around the helium core and ascend the RGB, and clump stars,
which have convection in the helium-burning core
\citep[e.g.][chap. 32]{1990sse..book.....K}.
\cite{2011arXiv1106.5946C}  has explained the higher values of
$\Tg$ in red-clump stars by the fact that g modes are excluded
from this convective core. Table~\ref{recap} summarizes the
properties of the stars presented in this paper.

Stars with different evolutionary states are clearly located in
different regions of the $\Dnu$ -- $\Tg$ diagram
(Fig.~\ref{spacings}). Compared to the bumped spacings, the
gravity spacings show a lower dispersion, especially for the
red-clump stars. If we set the mass limit between the first and
the secondary clump at 1.8\,$M_\odot$, the mean value of $\Tg$ in
the clump is $297\pm 23$\,s.  The secondary clump stars show a
broader distribution, with values in the range [150, 300\,s]. For
RGB stars, $\Tg$, in the range [60, 120\,s], has a clear
relationship with $\Dnu$. We note a slight decrease of $\Tg$
during the ascent of the first part of the RGB (for $\Dnu$
decreasing down from 40 to 7\,$\mu$Hz, equivalent to a radius
increase from about 2.5 to about 8\,$R_\odot$). This evolution
corresponds to the contraction of the core when the star ascends
the RGB. The tight $\Dnu$ -- $\Tg$ relation indicates that the
expansion of the envelope is closely related to the contraction of
the core.

We compared the observed $\Dnu$ -- $\Tg$ relation with the same
quantities obtained from a grid of stellar models with masses of
1, 1.2, and $1.4\, M_\odot$, from the zero age main sequence
(ZAMS) to the tip of the RGB. This range of mass corresponds to
the observed masses derived from the seismic estimates. These
models were obtained using the stellar evolution code CESAM2k
\citep{2008Ap&SS.316...61M}, assuming a gray Eddington
approximation and the B\"{o}hm-Vitense mixing-length formalism for
convection with a mixing-length parameter $\alpha=1.6$. The
initial chemical composition follows \cite{2005ASPC..336...25A},
with a helium mass fraction of $0.2485$. The period spacing was
computing using Eq.~\ref{def_period}, while the large separation
was computed using the asymptotic description $\Delta \nu = (2
\int_{0}^{R} \diff r / c)^{-1}$, where $c$ is the sound speed.

Qualitatively similar evolution was already shown in modeling
results
\citep{2011Natur.471..608B,2011ApJ...743..161W,2011arXiv1106.5946C},
but without a direct comparison to $\Tg$. In Fig.~\ref{spacings},
we note a close agreement between observed and modeled values for
$\Dnu \ge 11\,\mu$Hz. The spread around the evolution tracks is
low, less than 2\,\%. However, we do not see in the observed data
the clear mass dependence present in the models. The larger spread
in $\Tg$ for RGB stars with $\Dnu$ less than 11\,$\mu$Hz is
unexplained and deserves more work. Various reasons have to be
investigated, such as stellar rotation, the influence of the
mixing length and of overshooting, or an inadequacy of the
asymptotic relation in some specific cases. It may also correspond
to the first dredge-up \citep{2011arXiv1106.5946C}, which seems to
appear too late in the evolution models.

Finally, we note the presence of five red giants with a large
separation around 4.2\,$\mu$Hz and a g-mode spacing around 250\,s,
in a region of the $\Dnu$ -- $\Tg$ diagram clearly distinct from
the red clump. These apparent outliers are most plausibly red
giants that have exhausted helium in their core and have begun the
ascent of the asymptotic giant branch.

The frequency resolution is not high enough to provide reliable
measurements of $\Tg$ in RGB stars with $\Dnu \le 4\,\mu$Hz.

\subsection{Agreement with the asymptotic description}

We have found that all oscillation patterns with very many \gmmode
s can be fitted with the asymptotic relation. The highest
precision is obtained in spectra showing the most \gmmode s. G-m
modes present and identified near the $\ell=0$ and 2 ridges allow
the most precise observations. On the other hand, the absence of
\gmmode s most often hampers the measurement of $\Tg$, or results
in a limited accuracy, due to the possible misidentification of
the period spacing. The number of accurate determinations of $\Tg$
is therefore limited by the number of stars with very many \gmmode
s (\nombre\ stars studied in this work).

\begin{figure}
\includegraphics[width=4.44cm]{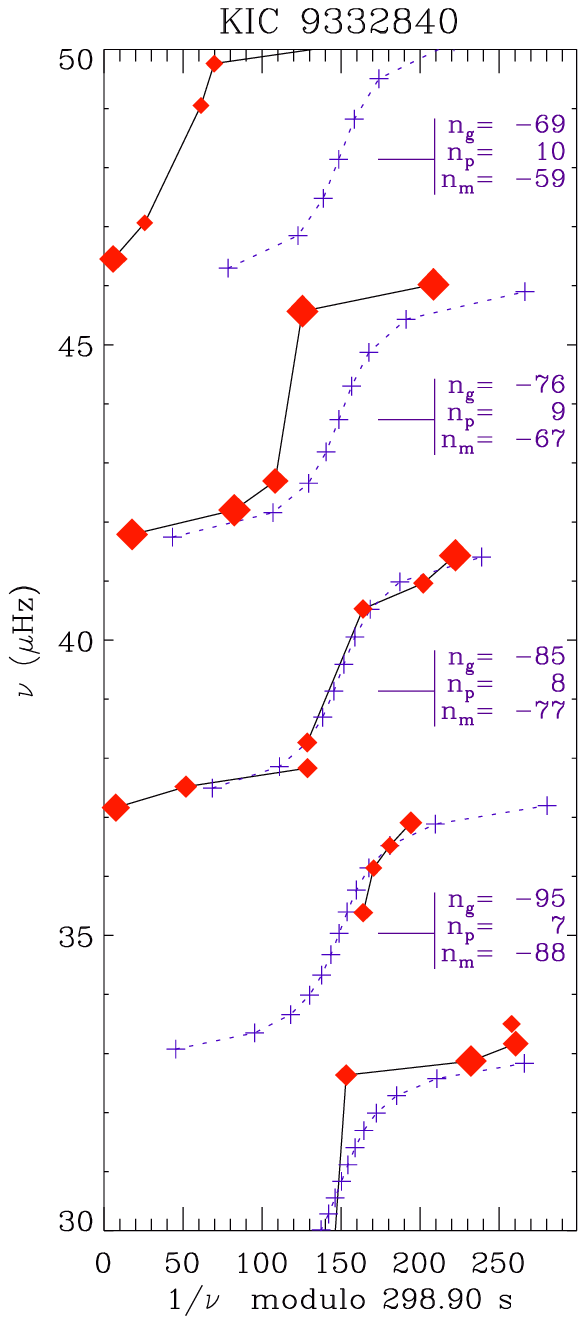}
\includegraphics[width=4.44cm]{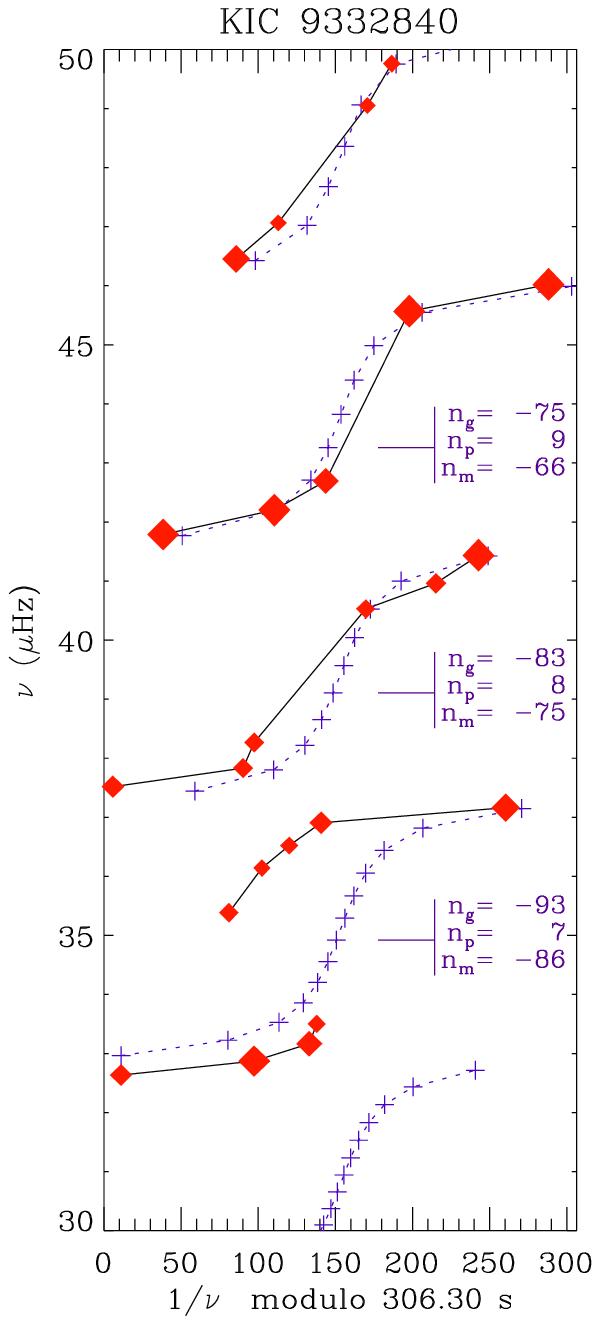}
\caption{Period \'echelle diagrams of the star KIC 9332840 based
on two different values of $\Tg$, presented as in
Fig.~\ref{gechelle}. {\sl Left:} the fit at low frequency requires
$\Tg = 298.9\pm 0.2$\,s. {\sl Right:} the high-frequency pattern
is better reproduced with $\Tg = 306.3\pm
0.2$\,s.\label{deuxfits}}
\end{figure}

A limited number of discrepant cases may occur. All these
correspond to oscillation spectra with relatively few peaks
extracted with the criterion defined in Sect.~\ref{gmmode}, and
never to a clear disagreement with the
asymptotic form. They fall mainly into two categories:\\
- Because dipole \gmmode s are observed everywhere in the
spectrum, possible errors may occur when they are located close to
other modes and may be mis-identified. The $\ell=2$ and $\ell=3$
non-radial modes may also present complex mixed-mode patterns, so
that an unambiguous identification of the mixed modes is not
possible. This occurs mostly at low frequency, when the resolution
limit is a problem. The increase of the observation time as
the \Kepler\ mission continues will solve these problems.\\
- In some cases, a gradient or a modulation of $\Tg$ possibly
explains the observations. These specific cases deserve additional
studies. Observationally, high signal-to-noise ratio spectra are
required for the analysis of small-amplitude mixed modes in a very
large frequency range. We show an example where a variation in
$\Tg$ of about 2.5\,\% is necessary to fit the lower and upper
part of the spectrum (Fig.~\ref{deuxfits}). According to the
propagation diagram, high-frequency g waves probe a less extended
cavity than low-frequency waves. This explains the higher value of
$\Tg$ measured at high frequency. Theoretically, the link of this
phenomenon with the shell structure of the red giant interior has
to be investigated in more detail.

Finally, a few remaining peaks not identified as dipole mixed
modes are likely to be $\ell=2$ or 3 mixed modes. Their study will
require additional work.

\subsection{Advantages of the asymptotic method}

\begin{figure}
\includegraphics[width=8.88cm]{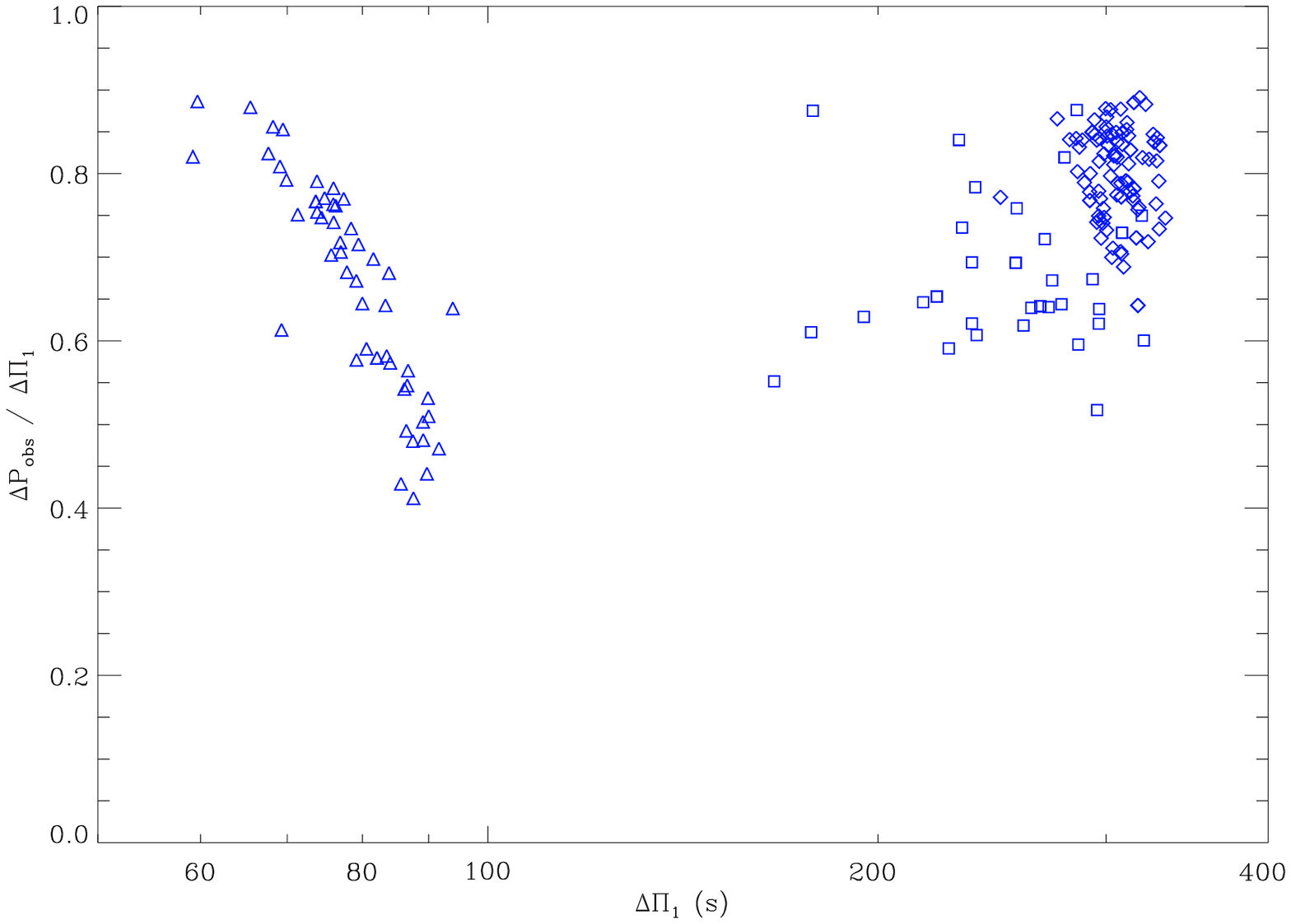}
\caption{Ratio $\tgobs / \Tg$  as a function of the period spacing
$\Tg$. Symbols have the same definition as in Fig.~\ref{spacings}.
\label{tgtobs}}
\end{figure}

The derived values of $\Tg$ can be compared to the values $\tgobs$
(Fig.~\ref{tgtobs}). The  ratio $\tgobs / \Tg$ depends on the
number of \gmmode s effectively used for deducing $\tgobs$. A low
ratio may occur because only few mixed modes close to the \pmmode
s are detected, or because the coupling is weak. Then, deriving
$\Tg$ from $\tgobs$ is possible but not precise. We also note that
rotational splittings (Sect.~\ref{rotation}) mimic low $\tgobs$
values, especially for RGB stars with large $\Dnu$.

The measurement of the period spacing is correlated with the
measurement of $\epsg$. At this stage, we have only determined
pairs of values ($\Tg$, $\epsg$). The uncertainties in $\Tg$
calculated with $\epsg = \epsgval$ are very small, typically about
0.2\,s for clump stars and 0.02\,s for RGB stars
(Table~\ref{recap}). Full uncertainties in $\Tg$ have to take into
account the unknown value of $\epsg$. They can be estimated for
the stars with the most \gmmode s. The uncertainties vary
inversely with the number of gravity nodes in the core: in
relative value, they are typically better than $1/(2\numax\Tg)$.
This represents about 0.5 to 1\,\% for clump stars, and between
0.3 and 1\,\% for RGB stars. We note that this high accuracy of
the measurement of $\Tg$ is based on the close agreement with the
asymptotic development. As shown by \cite{2011A&A...532A..86M},
the high absolute values of the gravity order $\ng$ ensure the
validity of the asymptotic relation and therefore strongly support
a close agreement. The coupling $\coup$ is also derived, with a
lower precision. Again, the uncertainties are much reduced when
\gmmode s far from the \pmmode s are observed.

The method is not only precise; it proves to be efficient, too.
For instance, \cite{2011Natur.471..608B} were able to identify the
g-mode spacing $\Tg$ in the star KIC 6928997 thanks to the
measurement of 21 mixed modes, with a Fourier spectrum performed
from a 13-month-long time series. We can now identify 45 mixed
modes, thanks to the longer observation run (22 months) and the
fit to the asymptotic relation (Fig.~\ref{gechelle}).

\begin{figure}
\includegraphics[width=8.88cm]{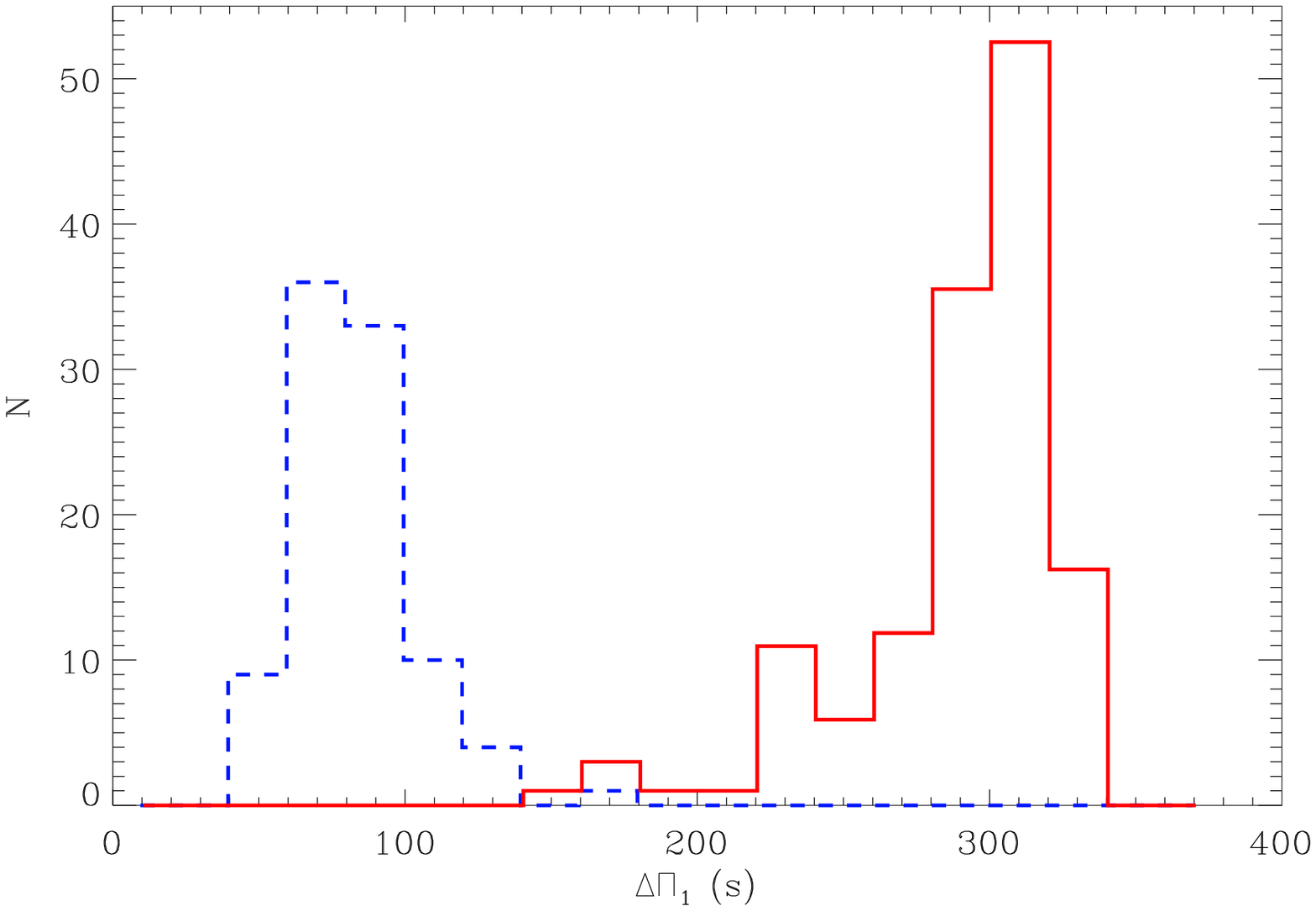}
\includegraphics[width=8.88cm]{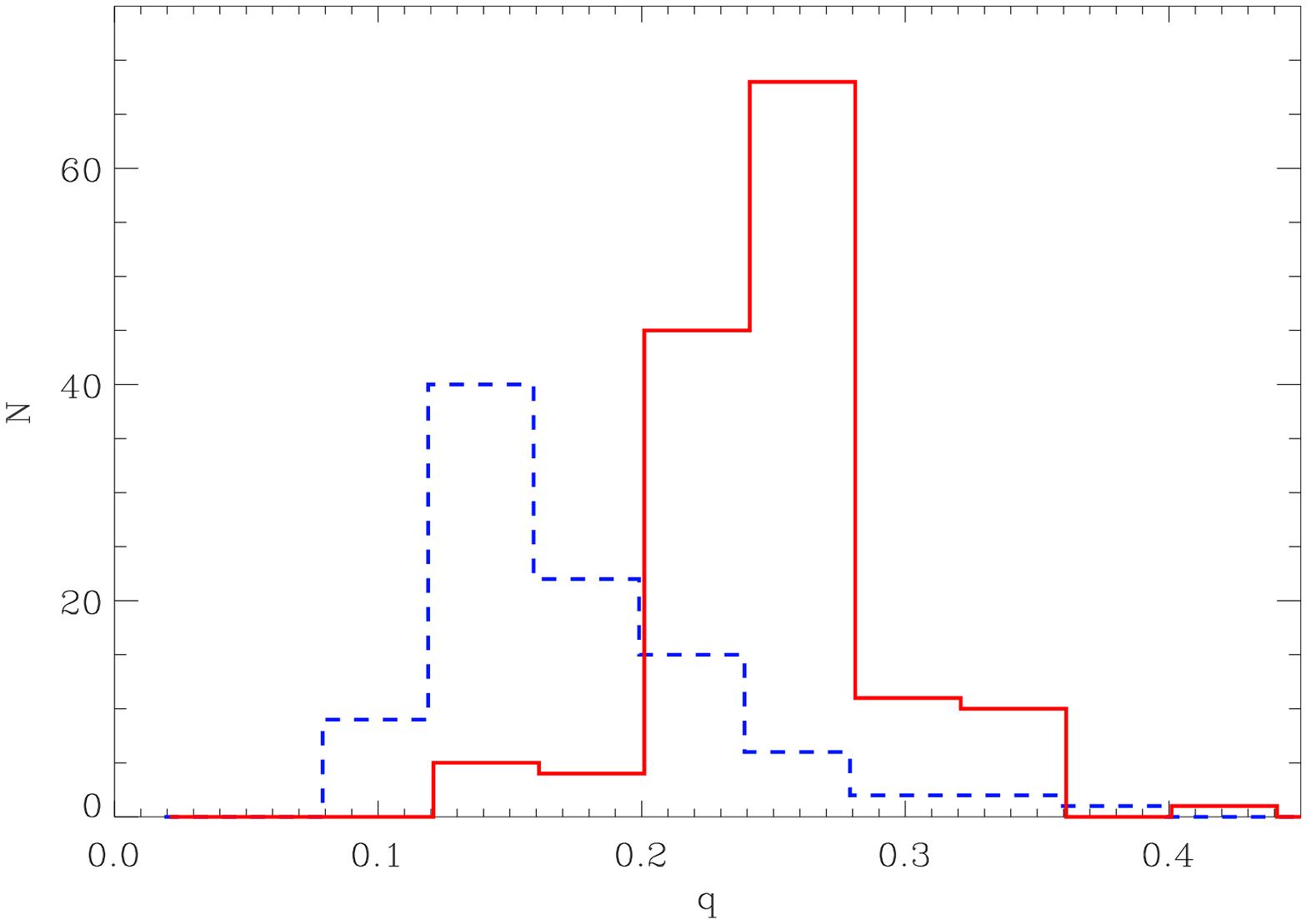}
\caption{{\sl Top:} histogram of $\Tg$, showing RGB stars (dashed
line) and clump stars (solid line). {\sl Bottom:} histogram of
$\coup$.\label{histo_Tg}}
\end{figure}

\begin{table*}
\caption{Characteristic parameters of the red giants named in the
paper}\label{recap}
\begin{tabular}{lrrccrrl}
 \hline
 KIC number & $\Dnu$ & $\numax^{(a)}$ & $\Tg^{(b)}$ & $\coup$ & $R\ind{seis}^{(c)}$ &
 $M\ind{seis}^{(c)}$ & evolutionary \\
            & ($\mu$Hz)&($\mu$Hz)& (s) &  & ($R_\odot$) & ($M_\odot$)&
            status$^{(d)}$
            \\
 \hline
 2013502 & 5.72 & 61.2 &  232.20$\pm$0.10  & 0.27$\pm$0.03 & 10.12 & 1.87 & 2nd clump\\
 3744043 & 9.90 &110.9 &   75.98$\pm$0.10  & 0.16$\pm$0.03 &  6.24 & 1.32 & RGB \\
 4044238 & 4.07 & 33.7 &  296.35$\pm$0.15  & 0.32$\pm$0.05 &       &      & clump\\
 5000307 & 4.74 & 42.2 &  323.70$\pm$0.30  & 0.26$\pm$0.06 & 10.36 & 1.38 & clump \\
 6928997 & 10.06&120.0 &   77.21$\pm$0.02  & 0.14$\pm$0.04 &  6.36 & 1.44 & RGB \\
 8378462 & 7.27 & 90.3 &  238.30$\pm$0.20  & 0.23$\pm$0.05 &  9.39 & 2.42 & 2nd clump\\
 9332840 & 4.39 & 41.4 &  298.9$\pm$0.2 - 306.3$\pm$0.2& 0.20$\pm$0.05 & 11.67 & 1.69 & clump\\
 9882316 &13.68 &179.3 &   80.58$\pm$0.02  & 0.15$\pm$0.05 &  5.41 & 1.63 & RGB \\
 \hline
\end{tabular}

Asymptotic mixed-mode parameters of  the red giant oscillation
spectra shown in the paper.

${}^{(a)}$: $\numax$ indicates the central frequency of the oscillation power excess.\\
${}^{(b)}$: this uncertainty assumes that $\epsg$ is fixed to 1/2.\\
${}^{(c)}$: $M\ind{seis}$ and $R\ind{seis}$ are the asteroseismic
estimates of the stellar mass and radius from scaling relations,
using $\Teff$ from the \Kepler\ Input Catalog
\citep{2011AJ....142..112B} .\\
${}^{(d)}$: the division between RGB and clump stars is derived
from Fig.~\ref{spacings}; the limit between the primary and
secondary (2nd clump) clump stars is set at 1.8\,$M_\odot$.\\
The complete table can be downloaded at the CDS via anonymous ftp
to \texttt{cdsarc.u-strasbg.fr (130.79.128.5)} or via
\texttt{http://cdsweb.u-strasbg.fr/cgi-bin/qcat?J/A+A/}

\end{table*}

\section{Discussion\label{discussion}}

\subsection{Overlapping modes\label{embedded}}

The close agreement with the asymptotic description for those
stars with very many mixed modes now allows us to extend the
mixed-mode identification to those stars with relatively few mixed
modes. Throughout, it is possible to identify the vast majority of
the peaks with a height five times above the mean background
level. This has also consequences for the degree identification.

\begin{figure*}
% 54 360 558 471
\includegraphics[width=17cm]{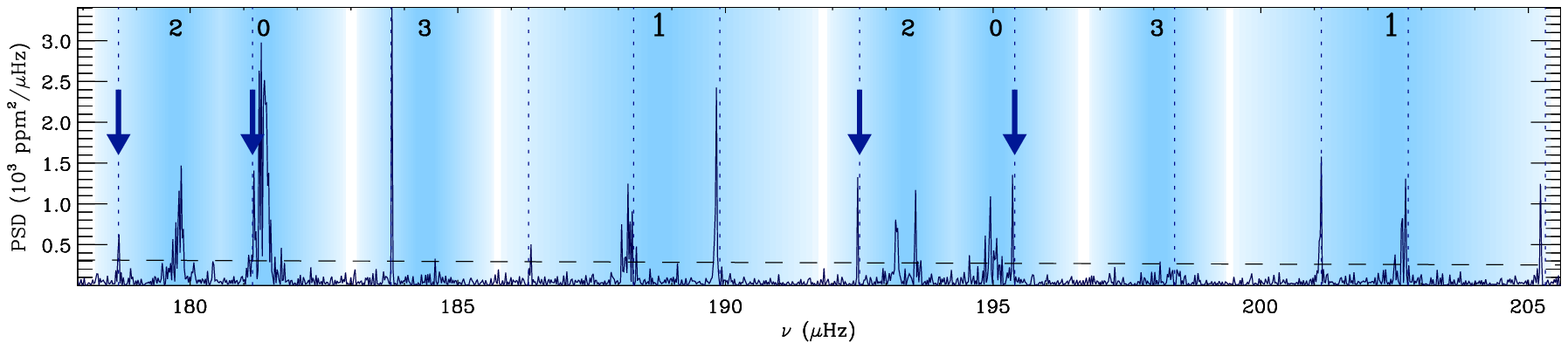}
\includegraphics[width=17cm]{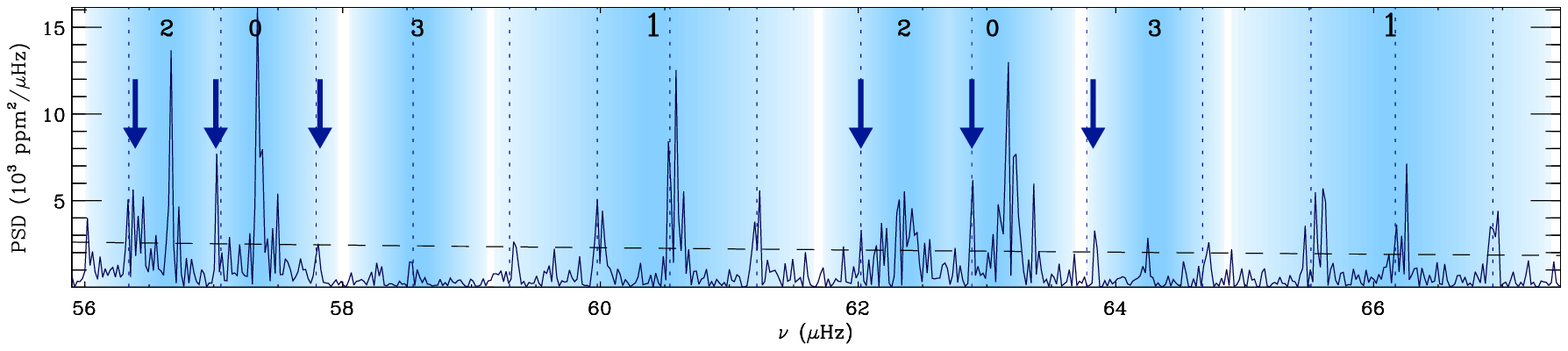}
\caption{Overlapping mixed modes identified a posteriori in the
power density spectra of three red giants. Vertical dotted lines
correspond to the dipole mixed-mode frequencies determined using
the asymptotic fit. Arrows indicate the modes assigned a
posteriori to $\ell=1$ mixed modes. The dashed lines show the
threshold level corresponding to eight times the stellar
background. {\sl Top}, KIC 9882316: additional peaks close to the
$\ell=2$ and 0 patterns present the characteristics of \gmmode s,
with a very narrow linewidth. {\sl Bottom}, KIC 2013502: the peaks
at 56.34, 57.06, 57.79, 62.02, 62.88, and 63.77\ $\mu$Hz, that is,
close to the uninterrupted series of \gmmode s, seem to be
$\ell=1$ mixed modes. This is not rare. \label{un}}
\end{figure*}

Dipole mixed modes can be present in the region of radial modes
and $\ell=2$ modes (Fig. \ref{un}). The narrow peaks corresponding
to very long-lived \gmmode s present in the vicinity of the
structure created by the short-lived radial p modes or complex
$\ell=2$ patterns are $\ell=1$ mixed modes. Coincidences are of
course possible. However, there would be so many coincidences that
their spurious existence is doubtful. Therefore, we argue that a
correct identification of the \gmmode\ pattern is definitely
necessary for addressing specific points such as the precise mode
identification or the measurement of the width of the radial modes
\citep{2011A&A...529A..84B}. We also note that the heights of
$\ell=1$ \gmmode s in the vicinity of $\ell=0$, 2, or 3 modes is
sometimes significantly boosted. This may result from
interference. We cannot exclude the possibility that the \gmmode\
heights are boosted by the energy of the other degree modes, even
if the orthonormality of the spherical harmonics of different
degrees seems to rule out mode coupling when oscillation
amplitudes are low, as is the case here. This deserves more
observations and simulations.

%\subsubsection{$\ell=3$ modes}

\begin{figure*}
% 54 360 558 471
\includegraphics[width=17cm]{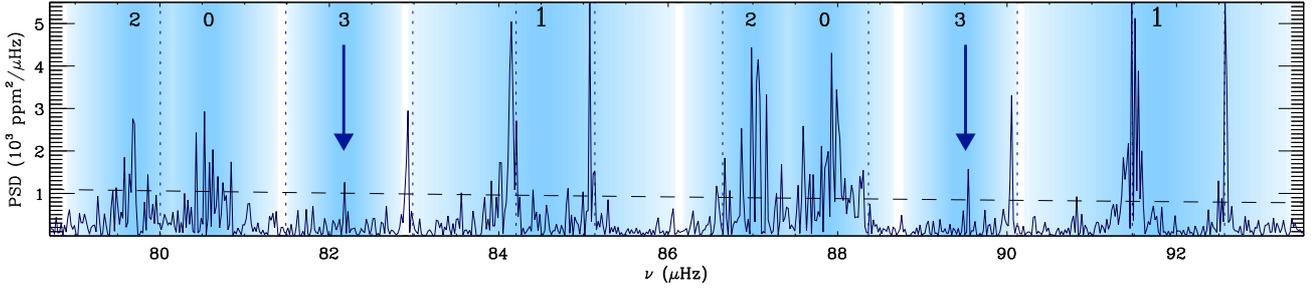}
\caption{Same as Fig. \ref{un}, but with arrows indicating the
modes a posteriori assigned to $\ell=3$ mixed modes. The identity
of these low-amplitude peaks is certain thanks to the prior
identification of $\ell=1$ mixed modes in the vicinity.
\label{trois}}
\end{figure*}

As already noted by \cite{2012A&A...537A..30M}, $\ell=1$ mixed
modes are present in the region of $\ell=3$ modes, so that
distinguishing true $\ell=3$ modes is difficult. Figure
\ref{trois} presents a typical case, with $\ell=3$ mixed modes
surrounded by $\ell=1$ mixed modes. We note the frequent
interferences between $\ell=1$ \gmmode s and $\ell=3$, and have to
conclude that the ensemble observations reporting the observations
of $\ell=3$ modes with high amplitudes may be overinterpreted
\citep{2010ApJ...713L.176B,2010ApJ...723.1607H,2012A&A...537A..30M}.
A significant number of peaks close to the expected location of
$\ell=3$ modes are most likely $\ell=1$ \gmmode s.

\subsection{Rotational splittings and mixed modes\label{rotation}}

\begin{figure*}
\includegraphics[width=16.88cm]{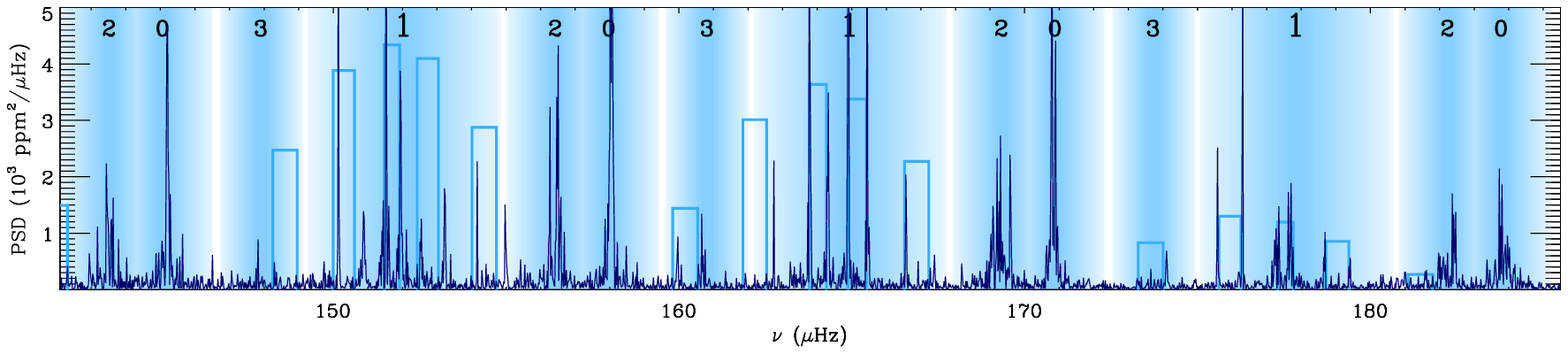}
\includegraphics[width=16.88cm]{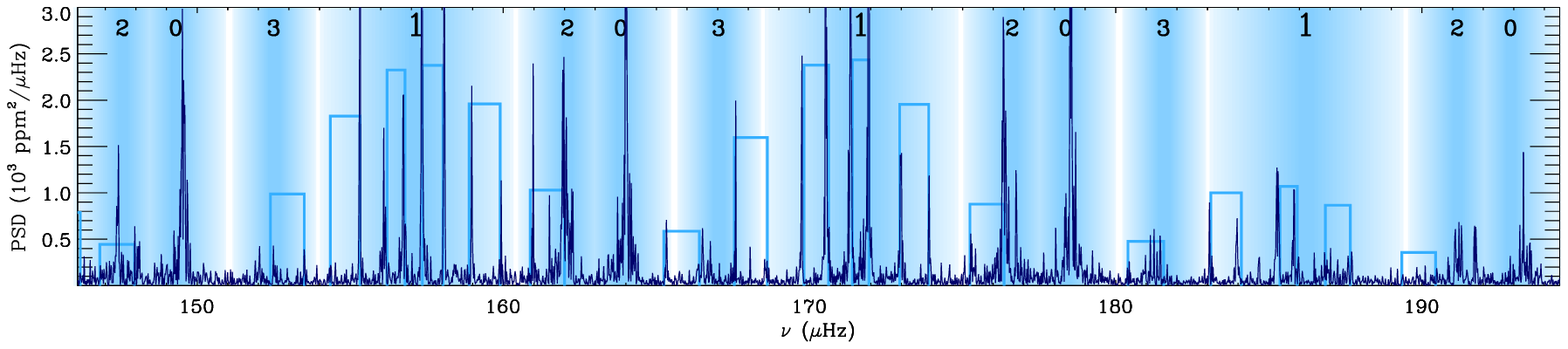}
\caption{Identification of rotational multiplets in two stars
observed with a nearly edge-on inclination and showing nearly
equal rotational splittings and mixed-mode spacing. Synthetic
$m=\pm 1 $ doublets based on the mixed-mode asymptotic relation
and on a simple modeling of the differential rotation are close to
the observed mixed modes. \label{mixterot}}
\end{figure*}

\cite{2012Natur.481...55B} have recently reported non-rigid
rotation in red giants, detected through a rotational splitting
that varies with frequency. In many cases, especially for stars
with $\Dnu$ above 10\,$\mu$Hz, the rotational splitting is so
close to the mixed-mode spacing that the direct identification of
the rotational multiplets is difficult. However, we showed that
the use of the asymptotic relation is able to provide the correct
identification of the rotational multiplets when coupled to a
simple ad-hoc model accounting for non-rigid rotation. We found
that an empirical modeling that takes into account the modulation
of the rotational splitting with a Lorentzian profile provides an
acceptable fit (Fig.~\ref{mixterot}). This profile $\Frot$
accounts for both the differential rotation observed by
\cite{2012Natur.481...55B} and the varying Ledoux coefficients
\citep{1951ApJ...114..373L}. Because the maximum observed
rotational splitting $\delta\nu\ind{rot,g}$ is small, $\Frot$
provides a splitting of the form
\begin{equation}\label{splitt}
  \nu (\nm,1,m) = \nu (\nm,1) + m \ \Frot (\nm,1)\ \dnugr,
\end{equation}
with $m$ being the azimuthal order. Locally, around a given dipole
pure p mode of radial order $\np$, for mixed-mode index $\nm$
associated via the coupling to this pressure radial order $\np$,
$\Frot$ can be expressed as
\begin{equation}\label{modulation}
 \Frot_{\np} (\nm, 1)
 =
 1 - {e \over 1 +  {\displaystyle{\left(
 { \nu (\nm, 1) -   \nu (\np, 1) \over \beta \Dnu}
   \right)^2}}} ,
\end{equation}
with $\nu (\np, 1)$ given by Eq.~\ref{tassoulp}. The constant
terms $e$ and $\beta$, empirically determined, are about 0.5 and
0.08, respectively, and $\dnugr$ is the maximum rotational
splitting observed for \gmmode s. The study of $\dnugr$ was not
conducted exhaustively and is beyond the scope of this paper.

\cite{2012Natur.481...55B} have noted that the heights of the
$m=+1$ and $-1$ modes of a given multiplet are very often not
symmetric, as observed in the solar case
\citep{2001MNRAS.327.1127C}. This emphasizes the interest of the
identification of the mixed-mode pattern for a correct
identification of the rotational structure of oscillation pattern.
Moreover, for similar reasons as those discussed in
Sect.~\ref{embedded} concerning overlapping modes, we stress that
in many cases the prior identification of the mixed-mode pattern
is essential to avoid misidentifications. In cases such as
presented in Fig.~\ref{mixterot}, despite a different behavior
with frequency of the rotational splittings (almost constant in
frequency apart from the periodic modulation due to $\Frot$) and
mixed-mode spacings (varying as $\nu^{-2}$), the near equality
between them greatly complicates the identification. For those
stars, the observed spacings $\tgobs$ were too small by a factor
of about 2, whereas the inferred $\Tg$ closely follow the trends
observed in Fig.~\ref{spacings}.

\subsection{Strength of the coupling }

Measuring the factor $\coup$ precisely is possible for about
75\,\% of the \nombre\ red giants. We note that the distribution
strongly depends on the evolutionary status of the red giants
(Fig.~\ref{histo_Tg}). The mean values are $\coup\ind{RGB} = 0.17
\pm 0.03$ and $\coup\ind{clump} = 0.25 \pm 0.05$. The lower values
for RGB stars indicate that the region between the g- and p-mode
cavities is larger than for clump stars. We also note the
existence of $q$ factors larger than 1/4, contradicting the
first-order modeling by \cite{1989nos..book.....U}. The full
interpretation of $\coup$ in terms of interior structure
properties requires more work. Indeed, the exact relation between
the factor $\coup$ and the properties of the region between the g
and p cavities remains to be established. The first-order
description given by \cite{1989nos..book.....U} is clearly not
sufficient to explain the values of this coupling constant for red
giants. In particular, for $\coup>1/4$, an extension of the Unno's
description will be necessary.

That $\coup$ is almost a fixed parameter at fixed evolutionary
status has two important consequences. When the quality of the
spectra does not allow the identification of \gmmode s with tiny
amplitudes, deriving a precise value of $\Tg$ remains possible,
assuming $\coup = \coupm$. When only mixed modes close to the
\pmmode s can be identified, it is still possible to obtain a
correct estimate of $\Tg$, thanks to a fit restricted to these
\pmmode s. In that case, the measurement is less precise, but
physically more significant than the value $\tgobs$ derived
between the \pmmode s.

\subsection{Mode heights of \gmmode s}

As previously noted, the observation of numerous \gmmode s so far
from the expected location of \pmmode s is a surprise.
\cite{2009A&A...506...57D} showed that for high-mass stars (2 to 3
$M_\odot$) significant heights are only expected near the \pmmode
s, in a frequency range limited to about 0.1-0.2\,$\Dnu$ around
them. However, most of the red giants observed by \Kepler\ have a
mass much below this range (Table~\ref{recap}). Among them, we
have identified spectra with \gmmode s observed almost everywhere.
However, there are also spectra with only a single \pmmode\ or a
very limited number of mixed modes per pressure radial order, as
predicted by \cite{2009A&A...506...57D}.

Theoretical information on the possible cause of this striking
variety  is limited at present. Clearly, we may imagine that the
coupling conditions between the two cavities may differ
significantly between the different red giants. However, we lack a
theoretical work similar to \cite{2009A&A...506...57D} for
low-mass stars in the range 0.8 to 2 $M_\odot$. A work like this
that would examine the propagation condition would help to
interpret the different observations and may give the key for
examining the deepest regions of the stars. These regions, at the
limit between the p- and g-mode cavities according to the seismic
view, correspond to the hydrogen-burning shell. Therefore, the
seismic observations allow us to investigate these regions, and to
test the different physical conditions, at the red giant bump,
before or after the first dredge-up.

We also recall that a population of red giants with very low
amplitude dipole modes has been reported by
\cite{2012A&A...537A..30M}. Unsurprisingly, the determination of
$\Tg$ for those stars is not possible. Theoretical work is
definitely necessary to understand the different mixed-mode
patterns.

\subsection{Pure gravity modes?}

Our analysis shows that in very many stars with \gmmode s, all
observed $\ell=1$ modes are mixed modes, with a significant
coupling between p and g waves according to the asymptotic
formalism. Their frequencies as well as their amplitudes are
strongly correlated with the pressure-mode pattern.
\cite{2012A&A...537A..30M} showed that the total visibility of
$\ell=1$ mixed modes surrounding a given p mode is consistent with
the expected visibility of the corresponding pure p mode,
emphasizing the role of the coupling with p modes in observing g
modes. This shows that the excitation of mixed modes is closely
related to the excitation mechanism of p modes provided by
turbulent convection in the uppermost layers of the stellar
envelopes.

From the observations, we deduce that we have not detected any
pure g modes, whose signature would have been a vertical line in
the period \'echelle diagram.  Pure g modes are expected to have
very low amplitudes at the stellar surface, which makes them hard
to detect. The prevalence of mixed modes is likely to be a
consequence of the high density in the red giant core, which
ensures that the \BV\ frequency is similar to or larger than the
p-mode frequencies, thus facilitating wave coupling. It is, of
course, difficult to draw many inferences on the core conditions
from stars that do not show a large number of g-m modes, except
perhaps that the conditions do not favor strong wave coupling
which, in itself, may be useful information.

\section{Conclusion\label{conclusion}}

The identification of red giant spectra with very many \gmmode s
and the use of the mixed-mode asymptotic relation proposed by
Goupil et al. (in prep.) allowed us to measure the gravity-mode
spacing $\Tg$. This provides a new and unique way to characterize
the physical conditions in the inner radiative regions of the red
giant cores.

We have observed in most cases a close agreement between the
observed mixed-mode spectra and the asymptotic relation. When this
agreement was not met, signal-to-noise ratio considerations were
sufficient to explain  the disagreement. Complex cases occur when
rotational splittings complicate the spectrum. However, a simple
modeling of the rotational splitting, accounting for differential
rotation, allowed us to separate the rotational splitting from the
mixed-mode spacing. The large variation in the height distribution
of the \gmmode s requires a study similar to that by
\cite{2009A&A...506...57D}, but extended to red giant stars with
masses in the range [0.8, 2\,$M_\odot$].

The identification of the $\ell=1$ \gmmode s allows the full
identification of all significant peaks in a red giant oscillation
spectrum. It also helps to perform the peak bagging more
efficiently. We stress that $\ell=1$ \gmmode s, present everywhere
in the oscillation spectrum, are often overlapping with other
degrees. In several cases, the previously reported $\ell=3$ modes
are, in fact, $\ell=1$ \gmmode s. The identification of the
complete $\ell=1$ mixed-mode pattern makes the spectrum clear.

Compared to the seismic constraints provided by the frequencies
$\Dnu$ and $\numax$, which are mainly sensitive to the average
stellar structure, the gravity period $\Tg$ directly probes the
core region. When available, the measurement of the gravity period
reaches a high level of accuracy, better than 1\,\%, thanks to the
many gravity nodes observed in the stellar core. A similar
accuracy is now necessary in red giant interior models.

We showed that the coupling factor and the heights of \gmmode s in
red giant oscillation spectra present a large variety of cases.
This is due to different physical conditions at the limit between
the helium core and the hydrogen envelope in the region where
hydrogen burns in shell, where the first dredge-up occurs.
Additional work will be carried out to take the full benefit of
these new asteroseismic constraints for probing these regions in
detail.

We showed that the contraction of the core in stars ascending the
RGB occurs with a tight relation between the period spacing and
the large separation. Asteroseismic measures provide clear
constraints on the red giant structure and evolution.

%______________________________________________________________
\begin{acknowledgements}
Funding for this Discovery mission is provided by NASA's Science
Mission Directorate. BM thanks Ana Palacios for meaningful
discussions about the red giant structure. YE acknowledges
financial support from the UK Science and Technology Facilities
Council. SH acknowledges financial support from the Netherlands
Organisation for Scientific Research (NWO). DS and TRB acknowledge
support by the Australian Research Council. JDR and TK acknowledge
support of the FWO-Flanders under project O6260 - G.0728.11. PGB
has received funding from the European Research Council under the
European Community's Seventh Framework Programme
(FP7/2007--2013)/ERC grant agreements n$^\circ$227224 PROSPERITY.

\end{acknowledgements}

\bibliographystyle{aa} % style aa.bst
\bibliography{biblio_g}

\end{document}